\documentclass[aps,prl,twocolumn,superscriptaddress,showpacs]{revtex4}

\usepackage{graphicx}
\usepackage{dcolumn}
\usepackage{bm}

\begin{document}

\topmargin       -0.2in


\title{Evidence for TeV Gamma-Ray Emission from the Galactic Plane}

\author{R.~Atkins}
\altaffiliation{Department of Physics, University of Utah, 115 South 
1400 East, Salt Lake City, UT 84112}
\affiliation{Department of Physics, University of Wisconsin, 1150 
University Ave, Madison, WI 53706}

\author{W.~Benbow}
\altaffiliation{Max-Planck-Institut f\"ur Kernphysik, Postfach 103980, 
D-69029 Heidelberg, Germany}
\affiliation{Santa Cruz Institute for Particle Physics, University of 
California, 1156 High Street, Santa Cruz, CA 95064}

\author{D.~Berley}
\author{E.~Blaufuss}
\affiliation{Department of Physics, University of Maryland, College
Park, MD 20742}

\author{D.~G.~Coyne}
\author{T.~DeYoung}
\affiliation{Department of Physics, University of Maryland, College
Park, MD 20742}

\author{B.~L.~Dingus}
\affiliation{Group P-23, Los Alamos National Laboratory, P.O. Box 1663, 
Los Alamos, NM 87545}

\author{D.~E.~Dorfan}
\affiliation{Santa Cruz Institute for Particle Physics, University of
California, 1156 High Street, Santa Cruz, CA 95064}

\author{R.~W.~Ellsworth}
\affiliation{Department of Physics and Astronomy, George Mason 
University, 4400 University Drive, Fairfax, VA 22030}

\author{L.~Fleysher}
\author{R.~Fleysher}
\affiliation{Department of Physics, New York University, 4 Washington 
Place, New York, NY 10003}

\author{G.~Gisler}
\affiliation{Group P-23, Los Alamos National Laboratory, P.O. Box 1663,
Los Alamos, NM 87545}

\author{M.~M.~Gonzalez}
\affiliation{Department of Physics, University of Wisconsin, 1150
University Ave, Madison, WI 53706}

\author{J.~A.~Goodman}
\affiliation{Department of Physics, University of Maryland, College
Park, MD 20742}

\author{T.~J.~Haines}
\affiliation{Group P-23, Los Alamos National Laboratory, P.O. Box 1663,
Los Alamos, NM 87545}

\author{E.~Hays}
\affiliation{Department of Physics, University of Maryland, College
Park, MD 20742}

\author{C.~M.~Hoffman}
\affiliation{Group P-23, Los Alamos National Laboratory, P.O. Box 1663,
Los Alamos, NM 87545}

\author{L.~A.~Kelley}
\affiliation{Santa Cruz Institute for Particle Physics, University of
California, 1156 High Street, Santa Cruz, CA 95064}

\author{C.~P.~Lansdell}
\affiliation{Department of Physics, University of Maryland, College
Park, MD 20742}

\author{J.~T.~Linnemann}
\affiliation{Department of Physics and Astronomy, Michigan State 
University, 3245 BioMedical Physical Sciences Building, East Lansing, MI 
48824}

\author{J.~E.~McEnery}
\altaffiliation{NASA Goddard Space Flight Ctr, Greenbelt, MD 20771}
\affiliation{Department of Physics, University of Wisconsin, 1150
University Ave, Madison, WI 53706}

\author{R.~S.~Miller}
\affiliation{Department of Physics, University of New Hampshire, Morse 
Hall, Durham, NH 03824}

\author{A.~I.~Mincer}
\affiliation{Department of Physics, New York University, 4 Washington
Place, New York, NY 10003}

\author{M.~F.~Morales}
\altaffiliation{Massachusetts Institute of Technology, Building 37-664H, 
77 Massachusetts Avenue, Cambridge, MA 02139}
\affiliation{Santa Cruz Institute for Particle Physics, University of
California, 1156 High Street, Santa Cruz, CA 95064}

\author{P.~Nemethy}
\affiliation{Department of Physics, New York University, 4 Washington
Place, New York, NY 10003}

\author{D.~Noyes}
\affiliation{Department of Physics, University of Maryland, College
Park, MD 20742}

\author{J.~M.~Ryan}
\affiliation{Department of Physics, University of New Hampshire, Morse
Hall, Durham, NH 03824}

\author{F.~W.~Samuelson}
\affiliation{Group P-23, Los Alamos National Laboratory, P.O. Box 1663,
Los Alamos, NM 87545}

\author{P.~M.~Saz~Parkinson}
\affiliation{Santa Cruz Institute for Particle Physics, University of
California, 1156 High Street, Santa Cruz, CA 95064}

\author{A.~Shoup}
\affiliation{Department of Physics and Astronomy, University 
of California, Irvine, CA 92697}

\author{G.~Sinnis}
\affiliation{Group P-23, Los Alamos National Laboratory, P.O. Box 1663,
Los Alamos, NM 87545}

\author{A.~J.~Smith}
\author{G.~W.~Sullivan}
\affiliation{Department of Physics, University of Maryland, College
Park, MD 20742}

\author{D.~A.~Williams}
\affiliation{Santa Cruz Institute for Particle Physics, University of
California, 1156 High Street, Santa Cruz, CA 95064}

\author{M.~E.~Wilson}
\affiliation{Department of Physics, University of Wisconsin, 1150
University Ave, Madison, WI 53706}

\author{X.~W.~Xu}
\affiliation{Group P-23, Los Alamos National Laboratory, P.O. Box 1663,
Los Alamos, NM 87545}

\author{G.~B.~Yodh}
\affiliation{Department of Physics and Astronomy, University
of California, Irvine, CA 92697}


\begin{abstract}
Gamma-ray  emission from a narrow band at the Galactic equator 
has previously been detected 
up to 30 GeV.  We report evidence for
a TeV gamma-ray signal from the Galactic plane by Milagro, a 
large field of view water Cherenkov detector for extensive air showers.
An excess with a significance of 4.5 standard deviations has been 
observed from the region of Galactic longitude
$l \in (40^{\circ},100^{\circ})$ and latitude $|b| < 5^{\circ}$. 
Under the assumption of a simple power law spectrum,
with no cutoff, in the EGRET-Milagro energy range, 
the measured integral flux is   
$\phi_{\gamma}(>3.5TeV) = (6.4 \pm 1.4 \pm 2.1) \cdot 10^{-11}$ $cm^{-2}s^{-1} sr^{-1}$.
This flux is consistent with an extrapolation of the EGRET
spectrum between 1 and 30 GeV in this Galactic region.   
\end{abstract}


\maketitle


Gamma rays are the best direct probe of cosmic rays outside the solar 
neighborhood. The interstellar medium, with its relatively high density in the 
Galactic plane, acts as a passive target for gamma-ray production by
energetic cosmic rays. Mechanisms include interactions
with  gas cloud nuclei that produce gamma rays via  $\pi^{0}$ decay,
as well as cosmic-ray electron bremsstrahlung and inverse Compton scattering
with the interstellar radiation field. Emission from a diffuse source 
concentrated in the narrow band along the Galactic equator was indeed 
detected by the space-borne detectors SAS 2, COS B
\citep{sas2_cos-b} and notably EGRET \citep{egret} at energies up to 30 GeV.
Above 1 GeV, the EGRET data in the region of the Galactic center 
show a hard spectrum $E^{-\alpha}$ with a differential spectral 
index $\alpha \approx 2.3$, and a flux enhancement of as 
much as 60\%, compared to models with $\pi^{0}$ production as the 
sole mechanism \citep{dermer_pio} and using the local cosmic ray spectrum.
Models that predict a Galactic flux 
enhanced by up to an order of magnitude over the $\pi^{0}$ mechanism at very 
high energies were proposed  
\citep{galactic_flux,galactic_flux2,galactic_flux3,strong_flux}.  
Upper limits have been set by several groups in the TeV range 
\citep{whipple,hegra_lim,tibet_lim} and above 180 TeV \citep{casa-mia}.

The Milagro Gamma Ray Observatory \citep{milagrito:nim,milagro_crab} is
a large-field-of-view telescope designed to detect gamma rays near 1 TeV
using water Cherenkov techniques to observe air shower particles that
survive to the ground level. It is located at a latitude of $36^{\circ}$
and an altitude of 2630 m in the Jemez Mountains, New Mexico, USA.  A
60m x 80m x 8m covered pond, filled with clear water, has a top layer of
450 photomultipliers (PMT), used to reconstruct the shower direction
with angular resolution of about $0.75^\circ$ from the relative PMT
timing.  A bottom layer of 273 PMT's is used for discrimination between
gamma-ray and the dominant hadron-induced air showers. A {\em
Compactness} cut, described in \citep{milagro_crab}, rejects about 90\%
of the hadronic background and retains about 45\% of the gamma-ray
signal for typical gamma-ray source spectra. On the opposite side of
this cut there are 9 times more cosmic rays and 1.2 times more gamma
rays, so that any gamma signal is suppressed by a factor of 7.4,
compared to the Compactness-cut signal. Additional cuts in this analysis
require a minimum of 50 top layer PMT hits, at least 20 of which
participate in the angle fit, a zenith angle of $\theta < 50^\circ$ and
a declination of $ 10^\circ < DEC < 60^\circ$. Periods with abnormal
event rate, zenith or azimuth distribution, corrupted records, or
unstable operation were excluded. The results presented are from three
calendar years of data collected by the Milagro detector starting July
2000. 

\begin{figure}[htbp!]
\centering
\includegraphics[width=7.5cm]{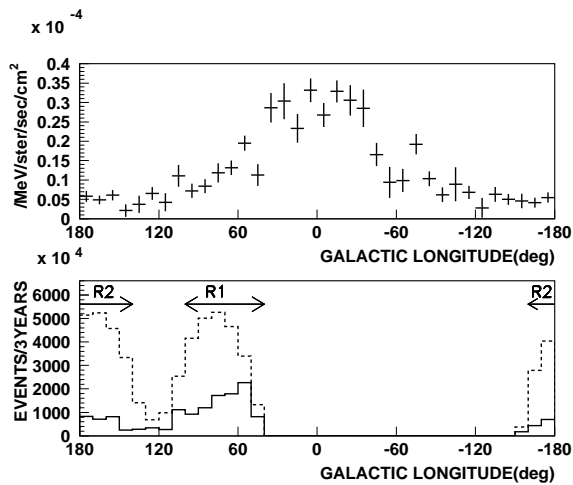}
\caption{Top: 4 to 30 GeV EGRET flux, Bottom: Milagro Exposure (dashed),
and expected relative significance (solid), along the Galactic equator. 
Regions R1 and R2 are indicated.}
\label{fig:results:signif_expect}
\end{figure}

Figure \ref{fig:results:signif_expect} shows the EGRET gamma-ray flux
along the Galactic equator\citep{{stan_hunter_private},{egret}}, peaking
near the Galactic center, and the Milagro exposure.  The Galactic center
is not visible to Milagro. Superposed is the nominal expected relative
significance, the product of the EGRET flux and the
square root of the Milagro exposure.  The region with highest expected
significance, R1 ($l\in (40^{\circ},100^{\circ})$), and a second region
of high Milagro exposure R2 ($l\in (140^{\circ},200^{\circ})$) both with
$|b|<5^{\circ}$ were selected {\em a priori}, as two critical regions for
separate statistical tests of a TeV gamma-ray signal. The Milagro 
exposures in R1 and R2 are similar. The average EGRET flux in R2 is about
a factor of two lower than in R1 and the extrapolation to TeV energies may well
be different in R1 and R2. R1 includes the Cygnus arm
and R2 is in the extreme outer galaxy. 

Gamma-ray emission from a gamma-ray source should appear as an excess of
the observed events ($N_{s}$) above the expected background count
($N_{b}$) in a candidate source direction bin. If the background is
isotropic in the celestial equatorial coordinate Right Ascension (RA)
and the detector acceptance is not changing over some time window, the
number of detected background events can be factorized into a
time-independent acceptance shape $G(x)$ in local angular coordinates
$x$ (e.g. Hour Angle and Declination) and a time-varying rate $R(t)$.
For a candidate source bin of interest (R1, R2, or any bin of Figure 2)
this gives $N_{b} = \int_{source\;bin} G(x) \cdot R(t) \; dx dt$. We
take advantage of a ``time-swapping'' method \citep{cygnus_methods} to
perform a Monte Carlo calculation of this integral.  The data is split
into 8-hour segments in which the $x$ and $t$ of recorded events are
considered random samples of G(x) and R(t). By pairing $x$ and $t$
randomly selected from these samples, new Galactic coordinates are
generated for background events.  $N_{b}$ is incremented whenever such
an event falls into the source bin.

We use a self-consistent modification, described in 
\citep{fleysher_meth}, of the method outlined above. The assumption that 
$G(x)$ is time independent is relaxed, allowing the incorporation of a 
small diurnal modulation, from atmospheric changes, that is observed in 
the zenith distribution of events. The statistical error of the 
background is calculated with ${\sigma_{N_{b}}}^{2} = A \cdot N_{b} + 
N_{b}/B$, where $A$ is the mean on-source to off-source exposure ratio 
and $B = 10$ is the number of time swaps per event \citep{errors}. The 
events arriving from the Galactic plane are excluded from the 
``time-swapping'', in order to maintain the statistical independence of 
$N_{s}$ and $N_{b}$ \citep{exclude}.

A very small anisotropy of the cosmic-ray background, at the level of a
few parts in $10^4$, was observed in a separate Milagro study
\citep{mil_anisotropy} and is applied as a correction in this analysis.
The anisotropy is consistent in shape and magnitude with reports from
several other experiments
\citep{hall_anis,kamiokande_anisotropy,tibet_anis}. Milagro's observed
anisotropy is described adequately by three longest wavelength harmonics
in right ascension (RA), whose amplitude and phase vary linearly with
declination (DEC). The anisotropy study provides a functional form for
the background correction in this analysis.  The subtracted sky map in
RA DEC coordinates, with R1 and R2 excluded, is fit to this form, then
extrapolated into these regions. The resulting fractional correction of
the background, applied in Table \ref{table:results:migsig}, is
$\delta_{bg}=-0.63 \pm 0.30 \cdot 10^{-4}$ and $\delta_{bg}=+0.04\pm
0.30 \cdot 10^{-4}$ for R1 and R2 respectively; the errors are the
statistical errors of the performed fit.

The test for a source signal is done taking R1 and R2 as two separate
single bins. The steps are given in Table \ref{table:results:migsig},
with the fractional excess defined by ${\cal F} = (N_{s}-N_{b})/N_{b}$. 
The statistical errors on $\cal F$ and $\delta_{bg}$ are combined in
quadrature in the last line.  An excess with a significance of 4.5
standard deviations is seen in R1 \citep{signif}, while no excess is
detected in R2.

\begin{table}[htbp!]
\begin{center}
\begin{tabular}{|c|c|c|} \hline
Region           & R1                            &  R2  \\ \hline
$N_{s}$ & $238,095,657$ $\pm 15430$  &  $254,800,416$ $\pm 15962$ \\ \hline
$N_{b}$          & $238,025,840 \pm 8003$ &  $254,826,272 \pm 8841$    \\ \hline
$N_{s}-N_{b}$    & $69,817 \pm 17,382$ & $-25,853\pm 18,247$ \\ \hline
${\cal F}_{raw}$ & $(2.93 \pm 0.73)\cdot 10^{-4}$ & $(-1.01 \pm 0.72)\cdot 10^{-4}$   \\ \hline
$\delta_{bg}$ & $(-0.63 \pm 0.30)\cdot 10^{-4}$ & $(+0.04 \pm 0.30)\cdot 10^{-4}$  \\ \hline 
$\cal F$  & $(3.56 \pm 0.79)\cdot 10^{-4}$ & $(-1.05 \pm 0.78)\cdot 10^{-4}$  \\ \hline
\end{tabular}
\end{center}
\caption {Calculation of the fractional 
excess ${\cal F} = (N_{s}-N_{b})/N_{b}$ for R1 and R2. $N_s$ is the 
observed event count; the error is on the mean of the underlying Poisson 
distribution.}
\label {table:results:migsig}
\end{table}

\begin{figure}[htbp!]
\centering
\includegraphics[width=8.5cm]{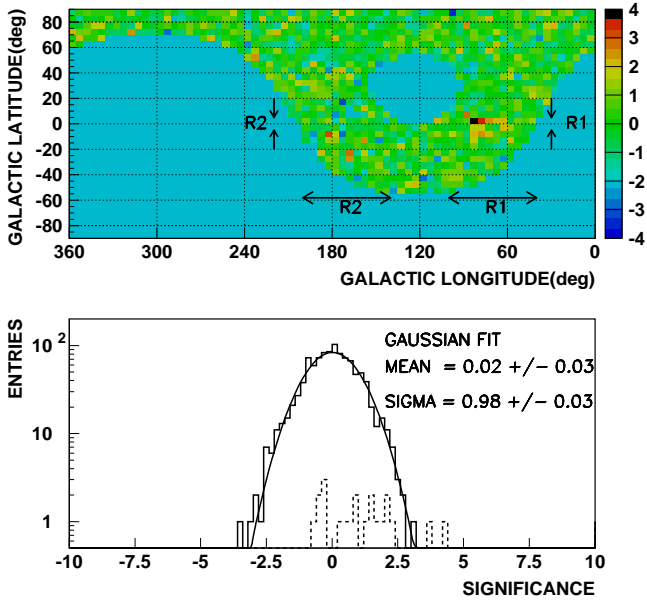}
\caption{Top (color online): Map with a color scale of the significance 
in Galactic coordinates in $5^{\circ}\times 5^{\circ}$ bins. Bottom:
Significance distribution outside R1(solid) and inside R1(dashed).}
\label{fig:results:significance_map}
\end{figure}

Figure \ref{fig:results:significance_map} (top) shows 
the two-dimensional color map of the significance
in Galactic coordinates. An enhancement in a ridge along 
the Galactic equator in R1 is the most prominent feature.
A Gaussian fit to the distribution of significances in
Figure \ref{fig:results:significance_map} (bottom)
has the requisite normal distribution with zero mean and unit variance,
while entries from R1 are shifted to the right.
 
\begin{figure}[htbp!]
\centering
\includegraphics[width=7.3cm]{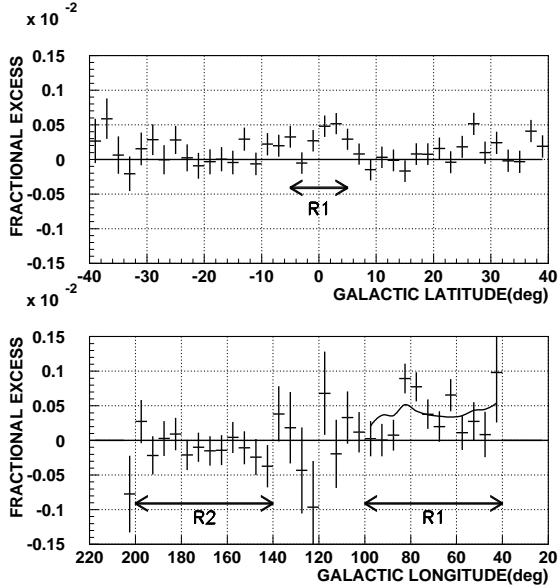}
\caption{Profiles of the fractional excess in latitude
for the R1 longitude band $l \in (40^{\circ},100^{\circ})$, and
in longitude for the latitude band $|b| < 5^{\circ}$ of 
R1 and R2. A fit of the EGRET longitudinal source shape 
(solid line) superposed in R1.}
\label{fig:results:profiles}
\end{figure}

Profiles of the fractional excess
${\cal F}$ in latitude and longitude
are shown in Figure \ref{fig:results:profiles}.      
The peak of the enhancement is just north of the Galactic equator
in Figures \ref{fig:results:significance_map} 
and \ref{fig:results:profiles}.
Such a shift  of the gamma signal     
towards positive latitudes was also seen in the EGRET data
\citep{egret}, interpreted as a large scale warp 
of the Galactic H 1 clouds \citep{warp}.

  To test for possible systematic problems in the analysis of the
R1 excess, an
identical analysis, including all modifications and corrections, 
was performed on 14 months of data, with the Compactness cut reversed to 
impoverish gamma rays relative to the cosmic-ray background. 
The result for R1, ${\cal F}^{\prime} = (0.68 \pm 0.39)\cdot 10^{-4}$   
agrees well with  
${\cal F}_{suppressed} = {\cal F}/7.4 = (0.48\pm 0.11) \cdot 10^{-4}$,  
that is expected in the absence of systematic effects.

We interpret the excess in R1 (Table \ref{table:results:migsig}) as
diffuse gamma-ray emission, or emission from unresolved gamma-ray
sources, or both. The ratio of integral gamma- and cosmic-ray fluxes is
calculated with:

\begin{equation}
{\cal R} = \frac{\phi_{\gamma}}{\phi_{(H+He)}} = 
   \frac{{\cal F}}{\eta(\alpha_{\gamma})}
\label{equation:flux_ratio}
\end{equation}

Because the intrinsic energy resolution of Milagro is limited,
individual event energies were not used in this analysis
\citep{energy_resolution}. Our flux determination assumes that the gamma
rays have a power law spectrum with no cutoff. The energy scale is
determined from air-shower and detector simulations and confirmed by
measurements of the Crab Nebula \citep{milagro_crab}. This gives a
median energy of approximately 3.5 TeV, with a 20\% systematic error,
for gamma rays from the Galactic equator. The coefficient
$\eta(\alpha_{\gamma})$ is the energy and transit averaged ratio of
gamma-ray to cosmic-ray effective area, with the Compactness cut,
obtained with Monte-Carlo air-shower and detector simulations
\citep{etavar}. We report results for the power law index connecting the
top EGRET point (10 - 30 GeV) and the Milagro point at 3.5 TeV,
$\alpha_{\gamma}=2.61$, for which $\eta = 6.2 \pm 2.0$. The error on
$\eta$ is the estimate of Monte-Carlo uncertainties, including the
energy scale error. We note that the flux is only sensitive to the
absolute energy scale of Milagro in proportion to the difference of
power law indices $E^{-(\alpha_{\gamma}- \alpha_{cr})}$.  The cosmic-ray
integral flux above 3.5 TeV/nucleus is $\phi_{H +He} = 1.2 \cdot 10^{-6}
cm^{-2}s^{-1} sr^{-1}$ \citep{jacee}.

\begin{figure}[htbp!]
\centering
\includegraphics[width=7.5cm]{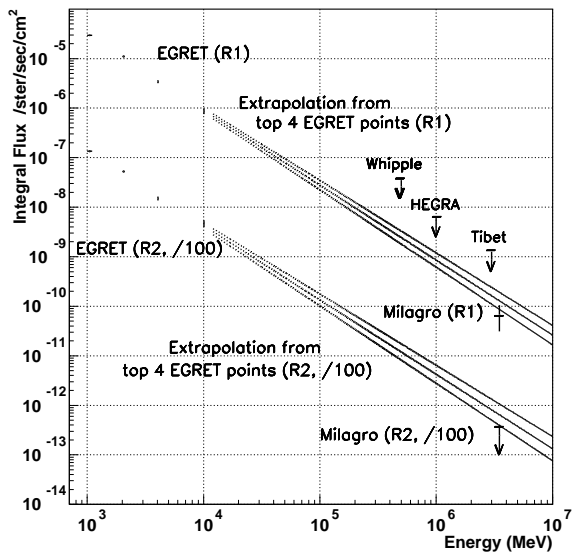}
\caption{Integral flux results of Milagro and EGRET, with $99\%$ c.l. 
         upper limits from Whipple ($l \in (38.5^{\circ}, 
         41.5^{\circ})$, $|b| < 2^{\circ}$) \citep{whipple}, 
         HEGRA ($l \in (38^{\circ}, 43^{\circ})$, $|b| < 5^{\circ}$) 
         \citep{hegra_lim}, Tibet ($l \in (20^{\circ}, 55^{\circ})$, 
         $|b| < 5^{\circ}$) \citep{tibet_lim}. The display for region R2 
         is shifted by $10^{-2}$.}
\label{fig:results:integflux}
\end{figure}

Integral flux results are shown in 
Table \ref{table:results:egmil_fit}, 
and plotted in Figure \ref{fig:results:integflux} together with the
EGRET data.

\begin{table} [htbp!]
\begin{center}
\begin{tabular}{|c|c|c|} \hline 
            & SINGLE BIN & MULTIBIN FIT  \\ \hline
${\cal R}$  & $(5.7 \pm 1.3 \pm1.8) \cdot 10^{-5}$ & $(5.7 \pm 1.3 \pm 1.8) \cdot 10^{-5}$   \\ \hline
$\phi_{\gamma}$ & $(6.8 \pm 1.5 \pm2.2) \cdot 10^{-11}$ & $(7.3 \pm 1.5\pm 2.2)\cdot 10^{-11}$ \\ \hline
$\alpha_{\gamma}$ & $ 2.61 \pm 0.03 \pm 0.05$ &  $ 2.60 \pm 0.03 \pm 0.05$  \\ \hline
\end{tabular} 
\end{center}
\caption{Ratio of the Milagro gamma-ray to cosmic ray flux, 
         integral flux above 3.5 TeV (in $cm^{-2} s^{-1} sr^{-1}$), 
         and the connecting EGRET-Milagro power law 
         index, for region R1. The last column uses a fit of the 
         longitude profile in Fig. \ref{fig:results:profiles} (bottom) 
         to the EGRET source shape. The results are not sensitive to 
         this choice.}
\label{table:results:egmil_fit}
\end{table}

For comparison, we fit the high end (1 GeV to 30 GeV) of the EGRET
spectrum \citep{egret,stan_hunter_private} in R1 to a power law,
obtaining the index $\alpha_{\gamma(EGRET)} = 2.51 \pm 0.05$, softer
than the Galactic center value of $\alpha \approx 2.3$, quoted earlier. 
Extrapolations of the EGRET fits in R1 and R2, with their 1 standard
deviation error corridors, are superposed on Figure
\ref{fig:results:integflux}.

As seen in Figure \ref{fig:results:integflux}, the emission from R1 at a 
3.5 TeV median energy is consistent with the extrapolation from the 
high-end EGRET data between 1 and 30 GeV. The absence of a detected 
excess in R2 requires a slight steepening of power law to 
$\alpha_{\gamma}(R2) > 2.66$ ($99\%$ cl) compared to 
$\alpha_{\gamma}(R1) = 2.61 \pm 0.03$. The Milagro flux upper limit of 
$\phi_{\gamma}(>3.5TeV) < 4 \cdot 10^{-11} cm^{-2} s^{-1} sr^{-1}$ 
($99\%$ cl) in R2 reaches the EGRET extrapolation for that region.

These findings do not require models predicting a strong enhancement
of the diffuse flux compared to conventional mechanisms, such as an
increased inverse Compton component \citep{pohl_esposito}, a harder
proton spectrum in the Galactic plane \citep{galactic_flux3}, or
contributions from unresolved supernova remnants \citep{galactic_flux2}. 
Assuming a power law spectrum for gamma rays with no cutoff below 10
TeV, we can exclude for R1 a hard spectrum with power law index
$\alpha_{\gamma} < 2.48$ ($99\%$ cl). The results are
consistent with a gamma ray power law index that asymptotically
approaches that of cosmic rays, as predicted if $\pi^0$ production
becomes the main source of the gamma flux \citep{dermer_pio}. 

    In summary, Milagro has observed a 4.5 standard deviation 
excess in the mid-longitude galactic plane region R1.   
The consistency of EGRET and Milagro data under the simple 
power law assumption reduces the motivation for speculation about 
more complicated gamma-ray spectra. With its one integral  
measurement, however, Milagro cannot rule out alternate models.
For the many possible multiparameter models, whose examination is 
beyond the scope of this letter, the Milagro 
result provides one constraint in their parameter space. In a 
model with a simple continuous power law spectrum from EGRET to Milagro 
energies, this measurement is evidence for gamma-ray emission at TeV energies,
with a flux that is consistent with an extrapolation 
from EGRET, and a median energy of about 3.5TeV for 
the detected gamma rays.

The authors thank Scott Delay and Michael Schneider for their dedicated
efforts on the Milagro experiment. We thank Stan Hunter for providing
the EGRET data. This work has been supported by the National Science
Foundation (under grants PHY-0075326, -0096256, -0097315, -0206656,
-0245143, -0245234, -0302000, and ATM-0002744) the US Department of
Energy (Office of High-Energy Physics), Los Alamos National Laboratory,
the University of California, and the Institute of Geophysics and
Planetary Physics.


\begin{thebibliography}{}

\bibitem{sas2_cos-b}
R.C. Hartman et al., Astrophys. J. {\bf 230}, 597 (1979).

\bibitem{egret}
S. D. Hunter et al., Astrophys. J. {\bf 481}, 205 (1997).

\bibitem{dermer_pio} 
C. D. Dermer, Astron. Astrophys. {\bf 162}, 223 (1986).

\bibitem{galactic_flux}
P. Chardonnet et al., Astrophys. J., {\bf 454}, 774 (1995).

\bibitem{galactic_flux2}
E. G. Berezhko and H. J. V\"{o}lk, Astrophys. J. {\bf 540}, 923 (2000).

\bibitem{galactic_flux3} 
F. A. Aharonian and A. M. Atoyan, astro-ph/0009009 (2000). 

\bibitem{strong_flux}
A. Strong et al., Astrophys. J. {\bf 613}, 956 (2004).

\bibitem{whipple}
S. LeBohec et al., Astrophys. J. {\bf 539}, 209 (2000).

\bibitem{hegra_lim}
F. Aharonian et al., Astron. Astrophys. {\bf 375}, 1008 (2001).

\bibitem{tibet_lim}
M. Amenomori et al., Astrophys. J. {\bf 580}, 887 (2002).

\bibitem{casa-mia}
A. Borione et al., Astrophys. J. {\bf 493}, 175 (1998).

\bibitem{milagrito:nim}
R. Atkins et al., Nucl. Instr. and Meth. {\bf A449} 478 (2000).

\bibitem{milagro_crab}
R. Atkins et al., Astrophys. J. {\bf 595}, 803 (2003).

\bibitem{stan_hunter_private}
The EGRET data was provided by S. D. Hunter, private communication (2002).

\bibitem{cygnus_methods}
D. E. Alexandreas et al., Nucl. Instr. and Meth. {\bf A328}, 570 (1993).

\bibitem{fleysher_meth}
R. Fleysher et al, Astrophys. J. {\bf 603}, 355 (2004).

\bibitem{errors} 
Calculated according to 
the statistic $U^\prime$, Eq.(8) of \citep{fleysher_meth}, a 
variant of the  Li and Ma statistic \citep{li_ma}  shown to converge
to it in \citep{fleysher_meth}.  

\bibitem{li_ma} 
T. Li and Y. Ma , Astrophys. J. {\bf 272}, 317 (1983). 

\bibitem{exclude}The slightly large $|b| < 7^{\circ}$ band
also excludes the possible
influence of the Crab Nebula, known to emit gamma rays in the TeV energy
region \citep{whip_crab,milagro_crab}.   
Disks of radius $5^\circ$ centered on the Sun and Moon, two moving
sinks of comic rays, were also excluded.

\bibitem{whip_crab}
T. C. Weekes et al., Astrophys. J. {\bf 342}, 379 (1989).

\bibitem{mil_anisotropy}
R. Atkins et al., in preparation.

\bibitem{hall_anis}
D. L. Hall  et al., J. of Geophys. Res., {\bf 104} A4, 6749 (1999).

\bibitem{kamiokande_anisotropy}
K. Munakata et al., Phys. Rev. {\bf D56}, 23 (1997).

\bibitem{tibet_anis} 
M. Amenomori et al. , Phys. Rev. Lett {\bf 93}, 061101 (2004). 

\bibitem{signif} An excess signal from this region was
also seen in two other Milagro analyses.

\bibitem{warp}
A. P. Henderson et al. Astrophys. J. {\bf 263}, 116 (1982). 

\bibitem{energy_resolution}
For more recent data, event energy reconstruction is
made feasible by locating the shower core with an ``outrigger''
array installed around the pond detector. 

\bibitem{etavar} $\eta$ varies by $\pm 10\%$ 
for $\alpha_{\gamma}$ between 2.5 and 2.9.

\bibitem{jacee} 
K. Asakimori et al., Astrophys. J. {\bf 502}, 278 (1998).
The error on this flux is negligible compared to 30\%. The 
contribution from nuclei heavier than H and He is negligible.

\bibitem{pohl_esposito}
M. Pohl and J. A. Esposito, Astrophys. Jour. {\bf 507}, 327 (1998).

\end{thebibliography}
\end{document}